# Critical current density for spin transfer torque switching with composite free layer structure

Chun-Yeol You

Department of Physics, Inha University, Incheon 402-751, Korea

Critical current density of composite free layer (CFL) in magnetic tunneling junction is investigated. CFL consists of two exchange coupled ferromagnetic layers, where the coupling is parallel or anti-parallel. Instability condition of the CFL under the spin transfer torque, which is related with critical current density, is obtained by analytic spin wave excitation model and confirmed by macro-spin Landau-Lifshitz-Gilbert equation. The critical current densities for the coupled two identical layers are investigated with various coupling strengths, and spin transfer torque efficiencies.

## 1. Introduction

The spin transfer torque magnetoresistive random access memory (STT-MRAM) offers superior performances such as non-volatility, scalability, speed, reliability, and power consumption compare to the conventional memories.[1] Spin transfer torque (STT) manifests itself by current induced magnetization switching (CIMS) in nanopillar magnetic tunneling junction (MTJ) structure.[2,3] Spin polarized electrons carried spin angular momenta from fixed layer to free layer, it causes free layer switching when the current density exceeds a critical value, critical current density $J_c$. Since the typical value of $J_c$ is over $10^{10}$ A/m$^2$, there are a lot of efforts have been made in the reduction of $J_c$.[4,5,6] The reduction of $J_c$ is one of the key issues in the research of STT-MRAM, because higher $J_c$ requires larger transistor size and causes serious Joule heating.[7] Another important issue is thermal stability of the free layer. Due to the scaling down of the free layer volume, the thermal stability factor, $E/k_BT$ (where $E$ and $k_BT$ are anisotropy and thermal energies of the free layer) becomes smaller. It causes degradation of the reliability of the STT-MRAM. Since the thermal stability is proportional to the volume of free layer, while $J_c$ is inversely proportional to the thickness of free layer, there is trade-off between them. In order to keep the thermal stability, while reducing $J_c$, alternatively synthetic ferrimagnetic (SyF) free layer structures have been proposed and tested.[8,9,10,11] In addition to SyF free layer structures, composite free layer (CFL) consisting of two ferromagnetic layers with various coupling types have been investigated.[12,13,14] However, surprisingly, there is no systematic theoretical approaches of the critical current density for the CFL structures including SyF, in spite of the $J_c$ of the single free layer is well studied.[15] In this study, we would like to propose an expression of the $J_c$ for CFL, the free layers consisting of two ferromagnetic layers with various kind couplings. We employed spin wave excitation model (SWM) to find instability conditions[16,17] of various CFL structure. The validities of the SWM are confirmed by macro-spin Landau-Lifshitz-Gilbert (MS-LLG) equation with STT contributions.

## 2. Spin wave excitation model

Let us consider MTJ stacks with fixed ferromagnetic layer ($F_{Fix}$), insulator layer (I), first ferromagnetic layer, ($F_1$), non-magnetic layer (NM), and second ferromagnetic layer ($F_2$) as shown in Fig. 1. The thickness of $F_1$ and $F_2$ layers are $d_1$ and $d_2$. Here we assumed that the resistance of I layer is much larger than other metallic layers, and the magnetization direction of $F_{Fix}$ layer is +*x* direction and rigid. The positive current means the electron flows from the fixed layer to free layer, the free layer prefers parallel configuration with fixed layer. Initially, the magnetization ($M_1$) of the $F_1$ layer is parallel to the –*x* direction, while the magnetization ($M_2$) of $F_2$ layer is aligned to +*x* (-*x*) direction for anti-parallel (parallel) coupling. The LLG equations with STT term for $F_1$ and $F_2$ layers are

$$\frac{d\vec{M}_1}{dt} = -\gamma\left(\vec{M}_1 \times \vec{H}_{eff}^1\right) + \frac{\alpha_1}{M_1}\left(\vec{M}_1 \times \frac{d\vec{M}_1}{dt}\right) + \mathrm{STT}_1, \tag{1}$$

$$\begin{aligned}\mathrm{STT}_1 = &-\frac{\gamma a_1 J}{M_1}\left(\vec{M}_1 \times \left(\vec{M}_1 \times \vec{P}\right)\right) - \gamma\left(b_0 J + b_1 J^2\right)\left(\vec{M}_1 \times \vec{P}\right) \\ &-\frac{\gamma a_{2,1}(-J)}{M_1 M_2}\left(\vec{M}_1 \times \left(\vec{M}_1 \times \vec{M}_2\right)\right) - \frac{\gamma b_{2,1}}{M_2}\left(\vec{M}_1 \times \vec{M}_2\right),\end{aligned} \tag{2}$$

$$\frac{d\vec{M}_2}{dt} = -\gamma\left(\vec{M}_2 \times \vec{H}_{eff}^2\right) + \frac{\alpha_2}{M_2}\left(\vec{M}_2 \times \frac{d\vec{M}_2}{dt}\right) + \mathrm{STT}_2, \tag{3}$$

$$\mathrm{STT}_2 = -\frac{\gamma a_{2,2} J}{M_1 M_2}\left(\vec{M}_2 \times \left(\vec{M}_2 \times \vec{M}_1\right)\right) - \frac{\gamma b_{2,2}}{M_1}\left(\vec{M}_2 \times \vec{M}_1\right). \tag{4}$$

Here, $\vec{H}_{eff}^{1,2}$ are effective field in $F_1$ and $F_2$ layers including external, anisotropy,

demagnetization, and exchange fields. $\alpha_{1,2}, \gamma$, and $\vec{P} = (1,0,0)$ are Gilbert damping parameter for $F_1$, $F_2$, gyromagnetic ratio, and the unit vector of $F_{Fix}$ layer magnetization direction. The $STT_1$ are torques acting on $F_1$ layer due STT by the $F_{Fix}$ and $F_2$ layers. Here, $a_1$ is so called Slonczewski term from $F_{Fix}$ layer defined by $a_1 = \eta_p \hbar / 2e\mu_0 M_1 d_1$, where $\eta_p$, $e$, and $\mu_0$, are the spin torque efficiency of $F_{Fix}$ layer, electron charge, and permeability, respectively. And $b_0$ and $b_1$ are field like terms of linear and quadratic coefficients of $J$, current density. It must be noted that the field like term is comparable with Slonczewski term in MTJ, while it is small in metallic systems. It has been known that different contributions of Brillouin zone integral are the physical reasons of the big difference between metallic and tunneling systems.[18] Furthermore, the current density or bias voltage dependences of field like terms are still controversial.[19,20,21,22,23,24,25] Therefore, we assumed the field like term of $b_0 J + b_1 J^2$ for the generality. The $a_{2,1}$ term is the Slonczewski torque acting on $F_1$ layer due to the $F_2$ layer, and it is defined $a_{2,1} = \eta_2 \hbar / 2e\mu_0 M_1 d_1$, $\eta_2$ is an spin torque efficiency of $F_2$ layer. Here, it must be emphasized that the direction of current must be considered as a negative at $F_1$ layer, we need extra minus sign in the third term of Eq. (2). $b_{2,1}$ and $b_{2,2}$ are another field like term acting on $F_1$ and $F_2$ layers, or we may call it as an interlayer exchange coupling term between $F_1$ and $F_2$ layers.[26,27] Since we consider a few nanometer thick metallic NM layer, the $b_{2,1}$ and $b_{2,2}$ depends on the thickness of NM, and they can be negative (positive) for anti-parallel (parallel) coupling. Furthermore, they are almost independent on the $J$. The $STT_2$ is the torque acting on $F_2$ layer due to $F_1$ layer. Here, $a_{2,2} = \eta_1 \hbar / 2e\mu_0 M_2 d_2$, $\eta_1$ are spin torque efficient of $F_1$ layer. In this study, we ignored the angular dependence of spin torque coefficients for the simplicity. Even though the angular dependence is considered, the main results of this work do not changed, but the detail dynamics might be varied.

First, let us consider SWM for anti-parallel (parallel) coupled CFL cases. We assume that the initial magnetization configuration is $\overrightarrow{M}_1 = -M_1\hat{x}$ and $\overrightarrow{M}_2 = M_2\hat{x}$ (or $-M_2\hat{x}$) for anti-parallel (parallel) coupling. Since we considered macro-spin model, there is no exchange field. For more simplicity we ignored anisotropy, and other possible effective fields such as dipole couplings between $F_{Fix}$, $F_1$, and $F_2$ layers. Therefore, the remaining effective field is an external magnetic field and demagnetization field. The we can write the effective field $\overrightarrow{H}_{eff}^{1,2} = \overrightarrow{H}_{ext} - \overrightarrow{N}_{1,2} \bullet \overrightarrow{M}_{1,2}$, where $\overrightarrow{N}_{1,2} = \left(N_x^{1,2}, N_y^{1,2}, N_z^{1,2}\right)$ are demagnetization vectors for $F_1$ and $F_2$ layers, and $N_x < N_y << N_z \sim 1$ for typical free layer geometry. We also assume $\overrightarrow{H}_{ext} = H_{ext}\hat{x}$, ($H_{ext} > 0$). When we turn on the current density $J$, spin wave is excited and it induces non-zero $y$- and $z$-components in $F_1$ and $F_2$ layers. Let us define the excited non-zero $y$- and $z$-components, $m_{y,z}^{1,2}(t)$. We put $m_{y,z}^{1,2}(t)$ contributions to the Eqs. (1)~(4), and linearize them up to first order of $m_{y,z}^{1,2}(t)$ because they are supposed to be small. After linearized, we obtain four couple differential equations of $m_{y,z}^{1,2}(t)$. Let us put $m_{y,z}^{1,2}(t) = \overline{m}_{y,z}^{1,2}e^{kt}$ with the simple harmonic oscillation model. With the same procedures in Ref. [16,17], we can build up a 4×4 matrix from four couple differential equations for anti-parallel $A^{AP}$ (or parallel $A^{P}$) cases as follows:

$$A^{AP,P} = \begin{pmatrix} A_{11}^{AP,P} & A_{12}^{AP,P} \\ A_{21}^{AP,P} & A_{22}^{AP,P} \end{pmatrix} \tag{5}$$

Where,

$$A_{11}^{AP,P} = \begin{pmatrix} k - \gamma\left(a_1 \mp a_{2,1}\right)J & -k\alpha_1 + \gamma H_{12}^{AP,P} \\ k\alpha_1 - \gamma H_{21}^{AP,P} & k - \gamma\left(a_1 \mp a_{2,1}\right)J \end{pmatrix}, \tag{6}$$

$$A_{12}^{AP} = A_{12}^{P} = \begin{pmatrix} a_{2,2}\gamma J\left(M_1/M_2\right) & b_{2,2}\gamma\left(M_1/M_2\right) \\ -b_{2,2}\gamma\left(M_1/M_2\right) & a_{2,2}\gamma J\left(M_1/M_2\right) \end{pmatrix}, \tag{7}$$

$$A_{21}^{AP,P} = \begin{pmatrix} -a_{2,1}\gamma J\left(M_2/M_1\right) & \mp b_{2,1}\gamma\left(M_2/M_1\right) \\ \pm b_{2,1}\gamma\left(M_2/M_1\right) & -a_{2,1}\gamma J\left(M_2/M_1\right) \end{pmatrix}, \tag{8}$$

$$A_{22}^{AP,P} = \begin{pmatrix} k \mp \gamma a_{2,2}J & \pm k\alpha_2 + \gamma H_{34}^{AP,P} \\ \mp k\alpha_2 - \gamma H_{43}^{AP,P} & k \mp \gamma a_{2,2}J \end{pmatrix}, \tag{9}$$

Here,

$$H_{12}^{AP,P} = H_{ext} \pm b_{2,1} + \left(b_0 J + b_1 J^2\right) + \left(N_x^1 - N_z^1\right)M_1, \tag{10}$$

$$H_{21}^{AP,P} = H_{ext} \pm b_{2,1} + \left(b_0 J + b_1 J^2\right) + \left(N_x^1 - N_y^1\right)M_1, \tag{11}$$

$$H_{34}^{AP,P} = H_{ext} - b_{2,2} \pm \left(N_z^2 - N_x^2\right)M_2, \tag{12}$$

$$H_{43}^{AP,P} = H_{ext} - b_{2,2} \pm \left(N_y^2 - N_x^2\right)M_2. \tag{13}$$

In order to have solutions, the determinant of matrix $A$ must be zero, and it is the secular equation for the variable $k$. The physical meaning of $k$ is clear: The imaginary parts of $k$ are corresponding to the angular frequencies of the excited spin wave. And if the real part of $k$ is negative, the excited spin wave is damped. However, when the real part of $k$ is positive value, $m_{y,z}^{1,2}(t)$ will diverge. It implies the given solution is instable, and the switching from the initial state is occurred. Therefore, when the real value of $k$ is positive, the corresponding $J$ is the critical current density $J_c$. For the single free layer, the instability condition is easily expressed as a function of given parameters since it is a quadratic equation for the $k$.[16,17] However, unfortunately, the secular equation is 4-th order of $k$ in this problem, so that the form of general solution is untractable. Therefore, from now on, we will solve $\left|A(k)\right| = 0$ equation by numerically. The obtained solutions will be compared with the full numerical solutions of Eqs. (1) ~ (4), the macro-spin approaches.

Before discuss more details of CFL case, let us reduce the problem to more simple case, a single free layer. If we assume $N_z$ ~ 1, and ignore $F_2$ layer, only 2×2 matrix $A_{11}$ with zero $a_{2,1}$ has to be solved. In that case, the instability condition is easily found, $J_c \approx \frac{\alpha_1}{a_1}\left(-H_{ext} + \frac{M_1}{2}\right)$, which is well-known result.

## 3. Results and Discussions

Let us consider a typical CFL of $F_1(d_1)$/NM/$F_2(d_2)$ structure for more details. For simplicity identical $F_1$ and $F_2$ layers are examined with zero external magnetic field, $M_1 = M_2 = 1.1\times10^6$ A/m, $\alpha_1 = \alpha_2 = 0.01$, $H_{ext}$ =0, and $\eta_p$ = 0.7 are substituted to Eq. (5). Here, the dimensions of $F_1$ and $F_2$ are 100 × 50 × 2 $nm^3$, and the corresponding demagnetization factors are evaluated and used in our calculations. First, we consider the interlayer exchange coupling field of $b_{2,1} = b_{2,2} = -5.0\times10^4$ A/m, and ignore the $b_0$ and $b_1$ contribution. The effect of $b_0$ and $b_1$ are not small for large $J$, but $b_0$ and $b_1$ contributions around $J_c$ are not significant. They will change some detail dynamics, but not the overall trends. Fig. 2 (a) ~ (d) shows $m_y^1(t)$ with SWM results with MS-LLG solutions for various $J$ = 1.2, 1.5, 1.9, and 2.2 × $10^{11}$ A/$m^2$ with $\eta_{1,2}$ = 0.4. The agreements between SWM and MS-LLG are excellent. Here, it must be pointed out that even we obtained finite positive $k$ for $1.5 \times 10^{11}$ A/m$^2$ case, but the switching does not occurred. The actual $J_c$ is $1.9 \times 10^{11}$ A/m$^2$ in this example. Since too small $k$, the longer time is required for the switching. Furthermore we simulate without thermal effect, there is no thermal activated switching effect. In our calculation, we found that $k > k_0 = 5 \times 10^8$/s are better criteria for the switching rather than positive $k$. The $k_0$ value means the characteristic time scale of corresponding excited spin wave is order of $2 \times 10^{-9}$ s, and it is related with switching time.

Trajectories of dynamic motions of $\vec{M}_1$ (blue) and $\vec{M}_2$ (red) for $J = 2.2 \times 10^{11}$ A/m$^2$ are plotted in Fig. 3. At the initially point, $\vec{M}_1$ and $\vec{M}_2$ are anti-parallel and they start precessions with an increasing amplitude as shown in Fig. 2 (d). After many precessions, their trajectories overcome some critical values, and finally they are reversed. During the switching processes, $\vec{M}_1$ and $\vec{M}_2$ are always anti-parallel due to the strong anti-parallel coupling between them.

Figure 4 (a) shows $J_c$ as a function of $b_2$ for $\eta_p = 0.7$ and $\eta_{1,2} = 0.4$. The results of SWM and MS-LLG are depicted together. The overall trends are similar for two results. We also plotted 4-nm thick single layer cases for the comparison. The single layer thickness is the sum of $d_1$ and $d_2$ and they are marked within a green circle. For the parallel coupling ($b_2 > 0$), $J_c$ are almost same values with the single layer one. The reasons are well explained with analogy of the coupled identical pendula model. It both pendula are identical, no force is acting to the pendula during in-phase motion, so there is no change of resonance frequency.[28,29] And it must be independent on the coupling strength. The relation between resonance frequency and $J_c$ will be discussed later.

For the anti-parallel coupling case ($b_2 < 0$), we find that $J_c$ increases with negative $b_2$.

It must be pointed out that this result is opposite to the recent experimental reports,[9,10] where strong anti-parallel coupling samples showed lowest $J_c$. In order to reveal the physical origin of the $J_c$ dependences on $b_2$, we plot spin wave excited frequencies $f_{\mathrm{swe}}$ as a function of $b_2$ in Fig. 4 (b). The $f_{\mathrm{swe}}$ is the imaginary part of corresponding $k$, and it is related with the resonance frequency of the system when $J = 0$. Even for the non-zero $J$, $f_{\mathrm{swe}}$ is almost same to the resonance frequency of the system. Since we considered two layers system, there are two possible resonance frequencies.[30] Here, we focus our attention to only lowest frequency state, which one is more easily excited. As shown in Fig. 4 (a) and (b), there is strong correlation between $f_{\mathrm{swe}}$ and $J_c$. The relation between them can be speculated in the spin wave excitation concept. When spin polarized current excites spin wave, the spin start precession with its own resonance frequency. Since the excited spin wave energy is proportional to its frequency, the higher characteristic frequency requires higher excited energy. Therefore, the strong anti-parallel interlayer exchange coupling causes high resonance frequency of the system due to the higher effective field.[29,31] As a result, the $J_c$ increases with negative $b_2$. Therefore tailoring of $b_2$ will be important, it can be easily achieved by alloying of non-magnetic spacer layer.[32,33] General cases such as $d_1 \neq d_2$, and/or $M_1 \neq M_2$ must be examined to explore more details.

Let's consider the effect of STT efficiency coefficients $\eta_{1,2}$. With the same parameters, we repeated the calculations with $\eta_{1,2} = 0 \sim 0.7$ for various $b_2$. The results are depicted in Fig. 5 (a) and (b). The open symbols are SWM, and solid symbols represent MS-LLG. Surprisingly, the $J_c$ are almost independent on $\eta_{1,2}$, regardless of $b_2$. It implies that the switching of $F_2$ layer is mainly occurred by $b_2$, not by $a_{2,2}$, and the contribution from $a_{2,1}$ to the switching of $F_1$ is also small. The $f_{swe}$ are also plotted for the comparison, and the strong relations between them are confirmed at a glance.

## 4. Conclusions

We investigate the critical current density for spin transfer torque switching with composite free layers. Spin wave excited model are developed for CFL and examined. The validity is confirmed by macro-spin LLG method. Most simple cases, two identical ferromagnetic layers, are calculated and we find that the $J_c$ strongly depends on the strength of the interlayer exchange coupling between $F_1$ and $F_2$ for anti-parallel coupling. And the $J_c$ is always larger than single layer case, which is contradictory to the recent experimental reports.[9,10] It must be noted that our numerical results are obtained for the identical two layers, and the non-identical layers will provide more complex behaviors and it will be explored elsewhere.

Acknowledgements

This work was supported by Inha University research grant. The author thanks Prof. K.-J. Lee for his helpful discussions.

**Figure Captions**

Fig. 1 Schematic diagram of layered structure. With rigid fixed layer ($F_{Fix}$), we considered a CFL consists of two ferromagnetic layers ($F_1$ and $F_2$) separated by non-magnetic layer (NM). The direction of positive current defined by from free layer to fixed layer.

Fig. 2 Motion of $m_y$ with spin polarized current density $J$ = 1.2, 1.5, 1.9, and 2.2 × $10^{11}$ A/m$^2$ for (a) ~ (d). The red open circles represent MS-LLG solutions, and the blue solid lines are results of SWM.

Fig. 3 Trajectory of dynamic motions of $F_1$ and $F_2$ layer magnetization for $J$ = 2.2 × $10^{11}$ A/m$^2$. Initially, $\vec{M}_1$ and $\vec{M}_2$ are anti-parallel, and they keep the anti-parallel coupling during the switching procedure.

Fig. 4 (a) Critical current density $J_c$ as a function of $b_2$. The red open rectangles and blue open circles represent the results of SWM and MS-LLG, respectively. The single layer cases are also depicted within green circle. (b) Corresponding $f_{swe}$ from SWM are plotted.

Fig. 5 (a) Critical current density $J_c$ as a function of $\eta_{1,2}$ for various $b_2$. Solid symbols are results from MS-LLG, while open symbols stands for the SWM results. (b) Corresponding $f_{swe}$ from SWM are plotted.

Figure1

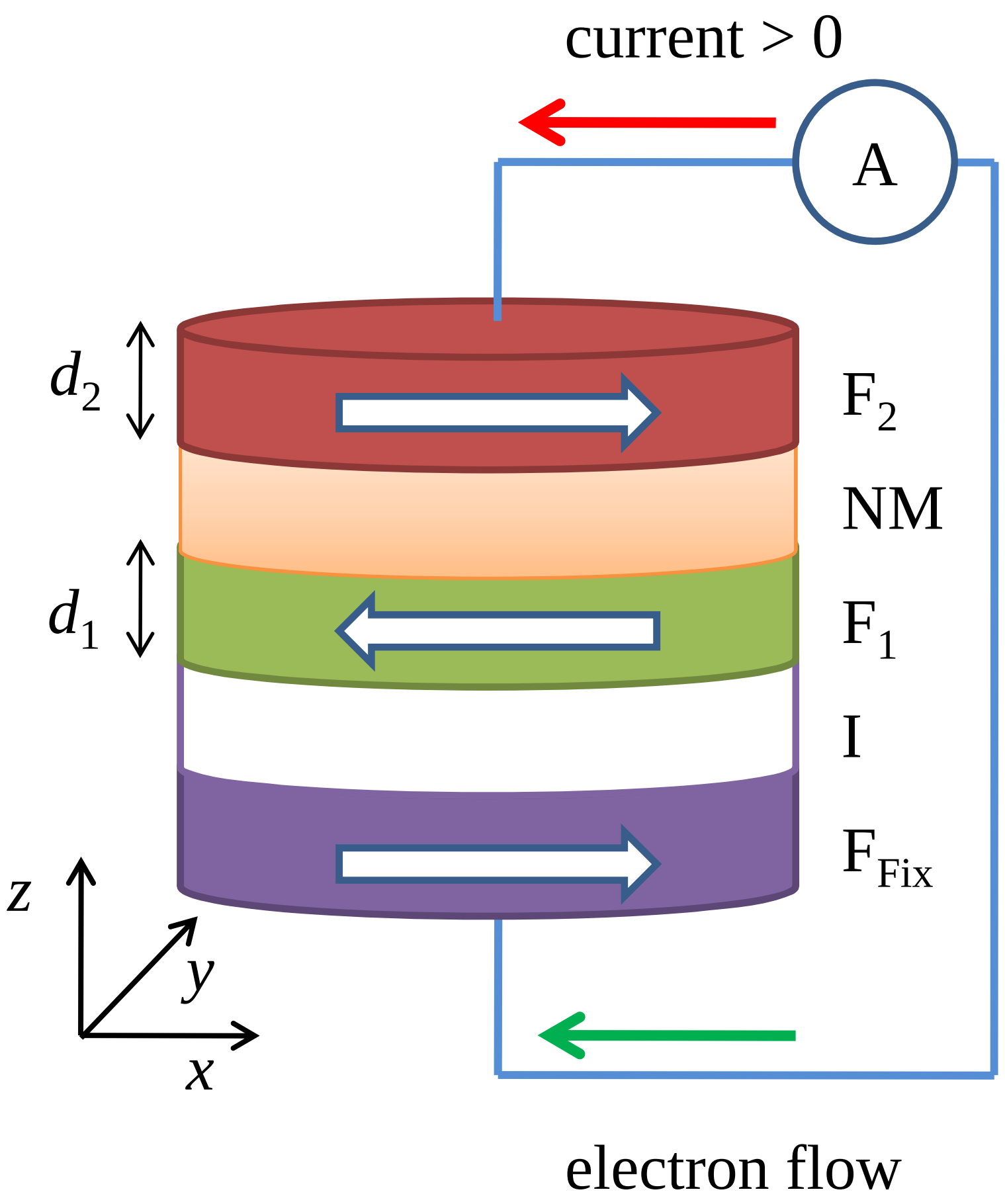

Fig. 1

Figure2

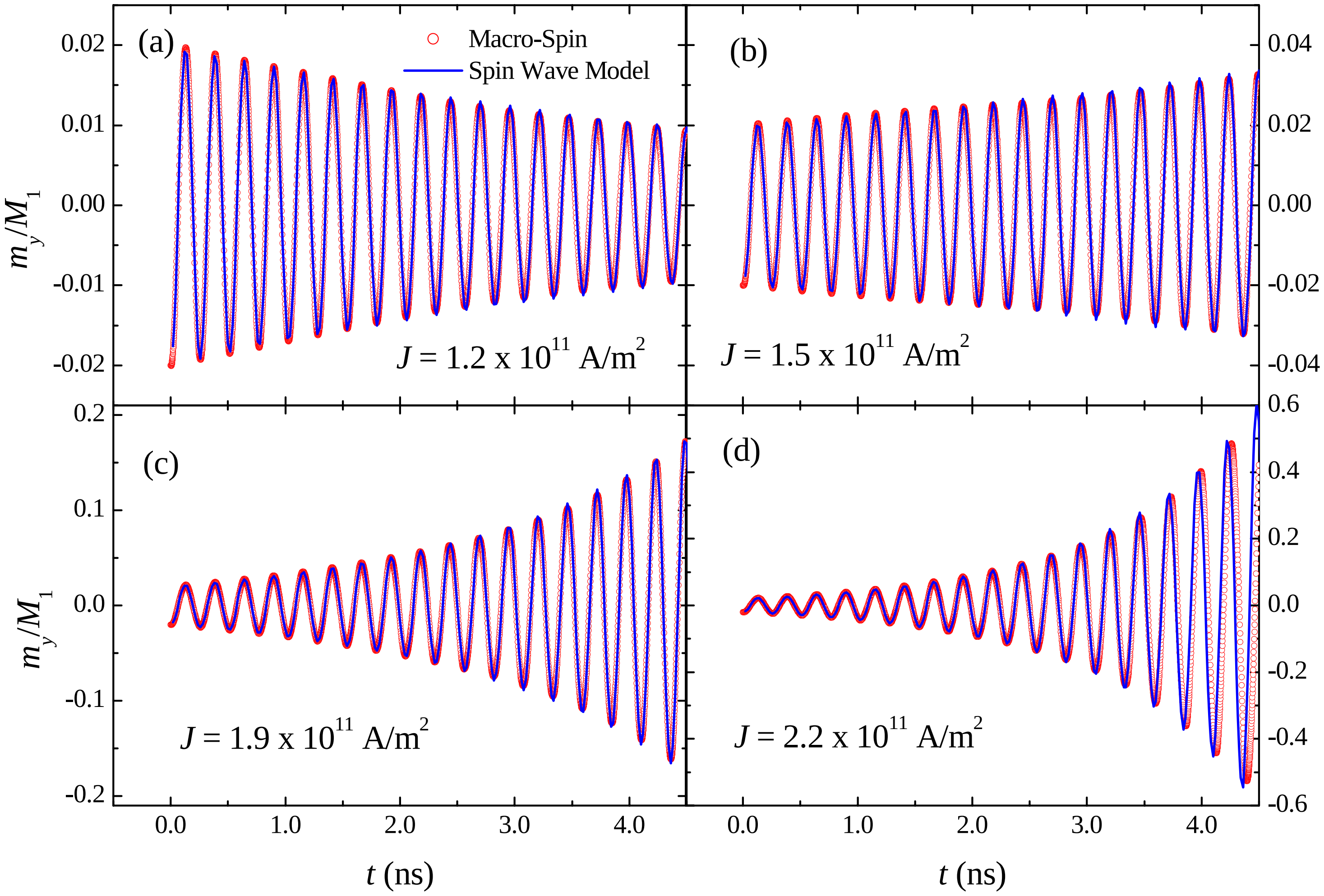

Fig. 2

Figure3

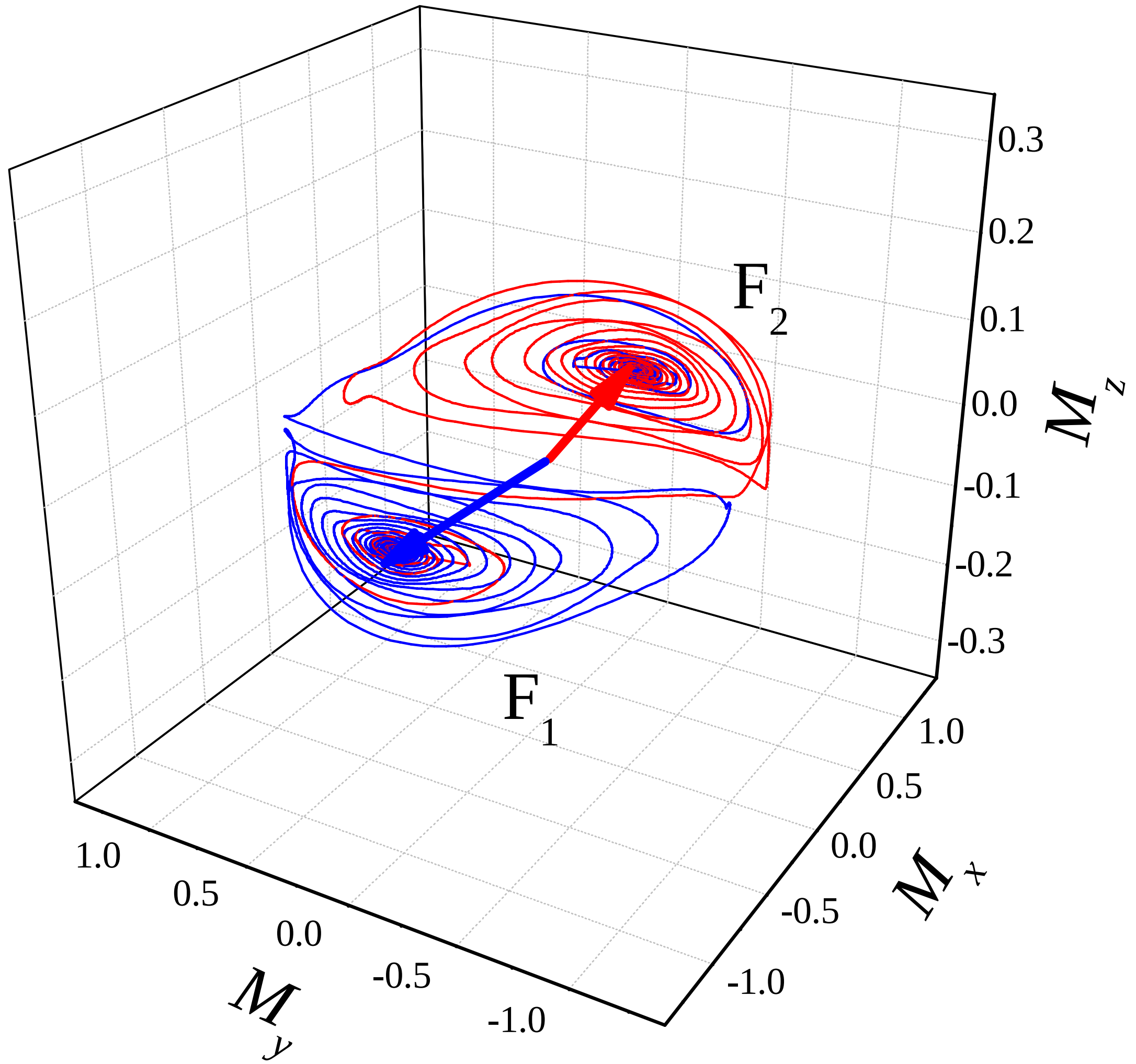

Fig. 3

Figure4

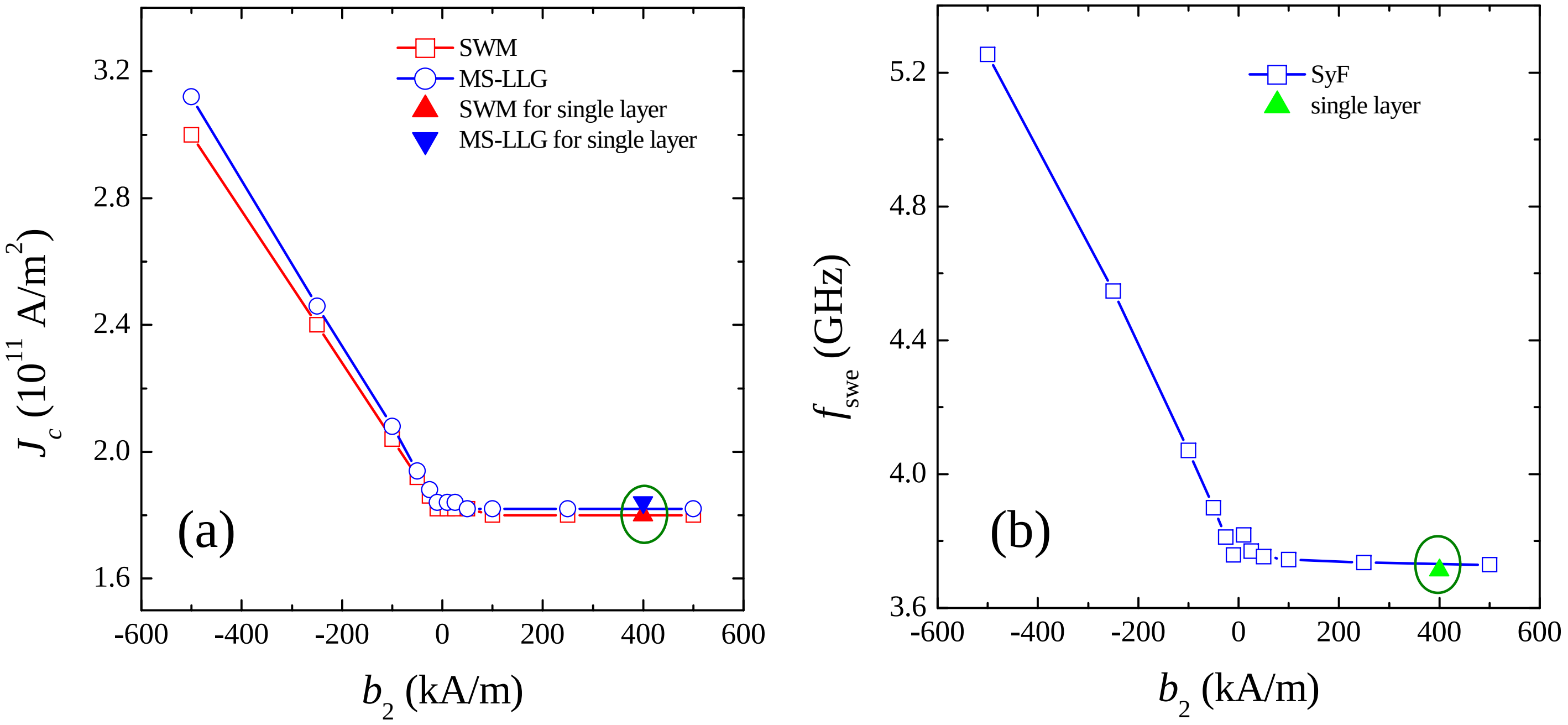

Fig. 4

Figure5

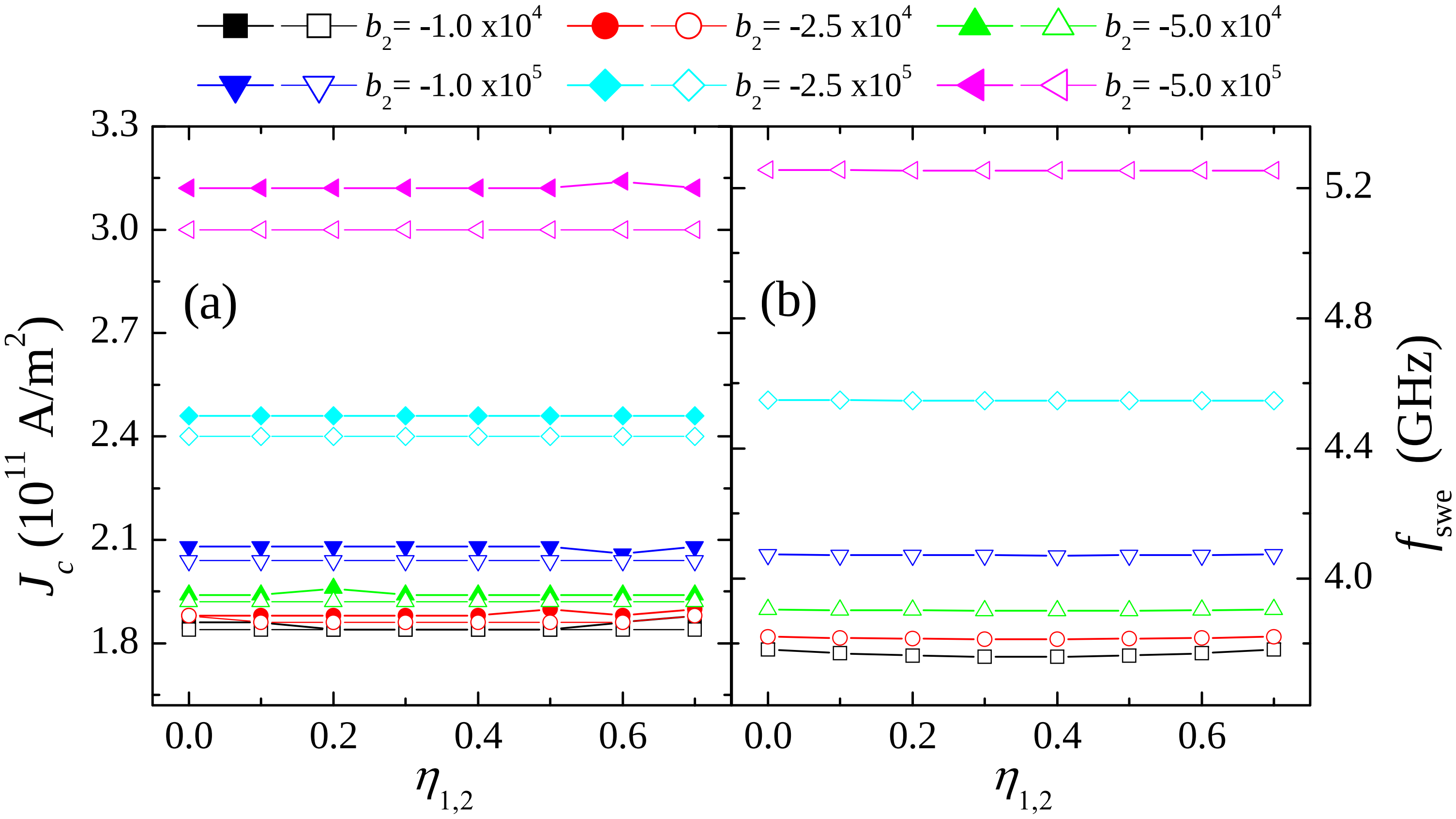

Fig. 5